\documentclass[12pt]{article}
\usepackage{amsmath}
\begin{document}
\begin{flushright}
KANAZAWA-11-04\\
\end{flushright}
\vspace*{1cm}

\begin{center}
{\Large\bf Thermal leptogenesis in a TeV scale model for neutrino masses}
\vspace*{1cm}

{\Large Daijiro Suematsu}\footnote{e-mail:~suematsu@hep.s.kanazawa-u.ac.jp}
\vspace*{1cm}\\

{\it Institute for Theoretical Physics, Kanazawa University, 
\\ Kanazawa 920-1192, Japan}
\end{center}
\vspace*{1.5cm} 

\noindent
{\Large\bf Abstract}\\
It is known that the radiative neutrino mass model proposed by Ma 
could be a consistent framework for dark matter, leptogenesis and 
suppressed lepton flavor violation if a neutral component of the 
inert doublet is identified as dark matter and the right-handed 
neutrinos are of $O(10^7)$ GeV or more. In the same model we explore 
another scenario such that right-handed neutrinos are in TeV regions 
and their lightest one is dark matter. It is shown that this scenario 
requires fine mass degeneracy to generate the appropriate baryon 
number asymmetry as in the case of resonant leptogenesis. As long as 
we impose the model to induce the baryon number asymmetry on the basis 
of thermal leptogenesis, we find that dark matter abundance can not be 
explained. If this scenario is adopted, the model has to 
be extended to include some new mechanism to explain it. 
\newpage
\section{Introduction}
The explanation of the origin of baryon number asymmetry in the universe 
is one of unsolved issues remained in the standard model (SM) \cite{bbn}.
Although various baryogenesis scenarios at high energy scales have 
been proposed, we now know that many of them do not work.
The generated baryon number asymmetry there is washed out by a sphaleron 
process in the electroweak interaction unless the $B-L$ symmetry is 
violated at some high energy scale \cite{sph}. 
Leptogenesis is a promising scenario since the $B-L$ symmetry is supposed 
to be violated through Majorana masses of right-handed neutrinos 
in the seesaw mechanism \cite{fy}.
As long as the right-handed neutrinos are heavy enough, the sufficient 
lepton number asymmetry can be produced through the out of thermal
equilibrium decay of the lightest right-handed neutrino \cite{leptg}.  
This lepton number asymmetry is processed to the baryon number 
asymmetry due to the sphaleron interaction.    
However, if we apply this scenario to supersymmetric models, 
the gravitino problem becomes serious \cite{gravitino}. 
Since reheating temperature required to escape the gravitino problem 
is too low to produce sufficiently heavy right-handed neutrinos 
in the thermal equilibrium, the required lepton number 
asymmetry could not be generated. A lot of models have been proposed to 
evade this difficulty \cite{dilimit,resonant,resonant1,soft,
sneut,nonth,others}.   

The radiative neutrino mass model proposed by Ma 
\cite{ma,raddm1,raddm2,raddm3} and also its supersymmetric 
extensions \cite{susyrad} are recognized as the models 
which can closely relate both small neutrino masses and 
the origin of dark matter (DM). 
Within the original Ma model, one can consider a simple scenario 
which simultaneously explains all baryon number asymmetry, 
correct dark matter abundance and realistic neutrino masses 
as discussed in \cite{ma1}. In that scenario, DM is identified 
with the lightest neutral component of an inert doublet \cite{inert} 
and the mass of the lightest right-handed neutrino is assumed to 
be of $O(10^{7})$ GeV or more.
On the other hand, if the right-handed neutrinos are assumed in TeV regions,
the same or worse situation discussed above is found in the model. 
Although the model seems to have several interesting features, 
the ordinary thermal leptogenesis seems not to work as the generation
mechanism of the lepton number asymmetry as long as DM is identified
with the lightest right-handed neutrinos.
An obstacle to it is just the lightness of the right-handed neutrinos,
while it brings interesting features to the model.

In that case nonthermal leptogenesis could be a consistent scenario 
for the baryon number asymmetry in this kind of model \cite{ad,deflm}. 
However, it is still an interesting subject to study in what situation 
the thermal leptogenesis could be applicable in this type of model.
For example, resonant leptogenesis might work also in this model 
if the right-handed neutrino masses are finely degenerate 
\cite{resonant,resonant1}.
In this scenario, resonant effect caused by the mass degeneracy 
enhances CP asymmetry in the decay of
the right-handed neutrino although light right-handed neutrinos 
require tiny neutrino Yukawa couplings. 
On the other hand, the washout brought by lepton number violating 
processes could be suppressed due to these small neutrino Yukawa couplings. 
As a result, the sufficient amount of lepton number asymmetry 
can be generated successfully there. 
Since the radiative neutrino mass model is characterized by the
neutrino mass generation mechanism different from the one considered in the
ordinary resonant leptogenesis, a new possibility for the 
thermal leptogenesis is expected to be found.
 
In this paper, we propose a scenario for thermal leptogenesis 
in a nonsupersymmetric radiative neutrino mass model.\footnote{A 
study for TeV scale leptogenesis in a different model can be 
found in \cite{hs}.}  
The scenario requires mass degeneracy for some fields, which is
realized only in this type of model.
It enhances the out of thermal equilibrium decay of a right-handed 
neutrino and also causes the Boltzmann suppression of washout 
processes of the lepton number symmetry. 
These features discriminate the scenario from the ordinary 
resonant leptogenesis. 
The scenario is expected to be applicable to a supersymmetric version
of the model straightforwardly. In that case the gravitino problem
could be escaped.
 
The paper is organized as follows. In section 2 we briefly address 
the model and assumptions imposed on the neutrino Yukawa couplings and
mass spectrum of new fields added to the SM. 
After that, we discuss the neutrino oscillation
parameters and the CP asymmetry obtained in the decay of a right-handed 
neutrino. We show that the sufficient amount of 
baryon number asymmetry could be generated in the model
through the study of Boltzmann equations relevant 
to the lepton number asymmetry.
In section 3 we address other phenomenological constraints on the 
model and discuss whether they could be consistent with the 
parameters required for the explanation of the baryon number asymmetry. 
We refer to the required modification for the DM scenario and also 
supersymmetric extension of the model.
Section 4 is devoted to the summary.
  
\section{CP asymmetry in a radiative seesaw model}
A model considered here is the radiative neutrino mass model 
studied in \cite{raddm1,raddm2}.
It is an extension of the standard model (SM) with a scalar doublet $\eta$ and 
three right-handed neutrinos $N_{R_i}$. 
They are assumed to have odd parity of a $Z_2$ symmetry, for which all SM 
ingredients have even parity.
The model is characterized by the following $Z_2$ invariant neutrino Yukawa 
couplings and scalar potential:
\begin{eqnarray}
-{\cal L}_y&=&h_{ij} \bar N_{R_i}\eta^\dagger\ell_{L_j}
+h_{ij}^\ast\bar\ell_{L_i}\eta N_{R_j}
+\frac{1}{2}\left(M_i\bar N_{R_i}N_{R_i}^c 
+M_i\bar N_{R_i}^cN_{R_i}\right), \nonumber \\
V&=&m_\phi^2\phi^\dagger\phi+m_\eta^2\eta^\dagger\eta
+\lambda_1(\phi^\dagger\phi)^2+\lambda_2(\eta^\dagger\eta)^2
+\lambda_3(\phi^\dagger\phi)(\eta^\dagger\eta) 
+\lambda_4(\eta^\dagger\phi)(\phi^\dagger\eta) \nonumber \\
&+&[\frac{\lambda_5}{2}(\phi^\dagger\eta)^2 +{\rm h.c.}],
\label{model}
\end{eqnarray}
where $\ell_{L_i}$ is a lepton doublet and $\phi$ is an ordinary Higgs doublet. 
Both Yukawa couplings of charged leptons and masses of 
the right-handed neutrinos are supposed to be real and flavor diagonal.
Since $\eta$ is assumed to have no vacuum expectation value, 
this $Z_2$ symmetry forbids to generate neutrino masses at tree level. 
Moreover, since the same $Z_2$ symmetry makes the lightest particle 
with its odd parity stable, it can be DM.
 
Neutrino masses are generated through one-loop diagrams. 
They can be expressed as\footnote{We assume
that $\lambda_5$ and $\langle\phi\rangle$ are real and positive.} 
\begin{equation}
{\cal M}^\nu_{ij}=\sum_{k=1}^3h_{ik}h_{jk}
\left[\frac{\lambda_5\langle\phi\rangle^2}
{8\pi^2M_k}\frac{M_k^2}{M_\eta^2-M_k^2}
\left(1+\frac{M_k^2}{M_\eta^2-M_k^2}\ln\frac{M_k^2}{M_\eta^2}\right)\right]
\equiv \sum_{k=1}^3h_{ik}h_{jk}\Lambda_k,
\label{nmass}
\end{equation}
where $M_\eta^2=m_\eta^2+(\lambda_3+\lambda_4)\langle\phi\rangle^2$.
In the following study, we consider flavor structure of 
the neutrino Yukawa couplings such as
\begin{equation}
h_{ei}^N=C\delta_{i2}h_i, \quad h_{\mu i}^N=h_{\tau i}^N\equiv h_i 
\quad (i=1,2); \qquad h_{e3}^N=h_{\mu 3}^N=-h_{\tau 3}^N\equiv h_3,
\label{yukawa}
\end{equation}
where a constant $C$ is supposed to be real, for simplicity.
In this case, the neutrino mass matrix is found to take a simple form as
\begin{equation}
{\cal M}^\nu=\left(
\begin{array}{ccc}
0 & 0 & 0\\ 0 & 1 & 1 \\ 0 & 1 & 1 \\ \end{array}\right)h_1^2\Lambda_1
+\left(
\begin{array}{ccc}
1 & 1 & -1\\ 1 & 1 & -1 \\ -1 & -1 & 1 \\ \end{array}\right) h_3^2\Lambda_3.
+\left(
\begin{array}{ccc}
C^2 & C & C\\ C & 1 & 1 \\ C & 1 & 1 \\ \end{array}\right)h_2^2\Lambda_2.
\label{nmass2}
\end{equation}
If $\lambda_5$ takes a small value such as $O(10^{-10})$, these mass 
eigenvalues can take suitable values for the explanation of neutrino
oscillation data \cite{oscil} even in the case where both $M_i$ 
and $M_\eta$ are of $O(1)$~TeV and $|h_i|=O(1)$.
This is the remarkable feature of this model.
The flavor structure defined by eq.~(\ref{yukawa}) 
is very interesting since it automatically induces the 
tri-bimaximal MNS matrix for $C=0$ such as \cite{raddm2}
\begin{equation}
U_{MNS}=\left(\begin{array}{ccc}
\frac{2}{\sqrt 6} & \frac{1}{\sqrt 3} & 0\\
 \frac{-1}{\sqrt 6} & \frac{1}{\sqrt 3} & \frac{1}{\sqrt 2}\\
\frac{1}{\sqrt 6} & \frac{-1}{\sqrt 3} & \frac{1}{\sqrt 2}\\
\end{array}\right)
\left(\begin{array}{ccc}
1 &0 & 0\\
0 & e^{i\alpha_1} & 0 \\
0 & 0 & e^{i\alpha_2} \\
\end{array}\right), 
\label{mns}
\end{equation}
where Majorana phases $\alpha_{1,2}$ are expressed as
\begin{equation}
\alpha_1=\varphi_3, \qquad
\alpha_2=\frac{1}{2}\tan^{-1}\left(\frac{|h_1|^2\Lambda_1\sin 2\varphi_1+
|h_2|^2\Lambda_2\sin 2\varphi_2}
{|h_1|^2\Lambda_1\cos 2\varphi_1+|h_2|^2\Lambda_2\cos 2\varphi_2}\right)
\label{cpphase}
\end{equation}
by using $\varphi_i={\rm arg}(h_i)$.

We find that the mass eigenvalues should satisfy
\begin{equation}
|h_1|^2\Lambda_1+|h_2|^2\Lambda_2\simeq 
\frac{\sqrt{\Delta m_{\rm atm}^2}}{2},  \qquad
|h_3|^2\Lambda_3\simeq \frac{\sqrt{\Delta m_{\rm sol}^2}}{3},
\label{c-oscil}
\end{equation}
where $\Delta m^2_{\rm atm}$ and $\Delta m^2_{\rm sol}$ stand for 
squared mass differences required by the neutrino oscillation data
\cite{oscil}. 
Recent T2K and Double Chooz data suggest a nonzero value for
$\theta_{13}$ \cite{t13}. 
We may include it in this model by introducing a nonzero
$C$ as a perturbation for eq.~(\ref{mns}).
In fact, if we assume that $C|h_2|^2\Lambda_2 \ll |h_1|^2\Lambda_1$ and
$\frac{C^2}{9}|h_2|^2\Lambda_2\ll |h_3|^2\Lambda_3$ are 
satisfied, the conditions (\ref{c-oscil}) for the neutrino masses are
good approximation even in the case with $C\not= 0$. 
If the model parameters $\lambda_5$, $M_\eta$ and $M_{1,2,3}$ are fixed 
to realize eq.~(\ref{c-oscil}), it is obvious that the neutrino oscillation 
data can be explained successfully in the model.

If we take account of lepton flavor violating processes such as 
$\mu\rightarrow e\gamma$, the experimental bounds might require $M_{1,2}<M_3$
\cite{raddm2}. Since the lightest $Z_2$ odd field is stable, either 
$N_{R_1}$ or a neutral component of $\eta$ could be a DM candidate. 
The latter possibility has been considered in a lot of articles
\cite{inert}. In that case the thermal leptogenesis could give appropriate
baryon number asymmetry consistently as long as $M_1>10^7$ GeV is
satisfied \cite{ma1}.
In the present study we adopt the former possibility and assume the 
following mass spectrum for the $Z_2$ odd fields: 
\begin{equation}
M_1~{^<_\sim}~M_{\eta}~{^<_\sim}~M_2<M_3.
\label{nspec}
\end{equation}

Neutrino Yukawa couplings are controlled by the constraints 
in eq.~(\ref{c-oscil}).
In Fig.~1 we plot the neutrino Yukawa couplings $|h_{1,3}|$ as a
function of $M_{1,3}$ by imposing the neutrino oscillation data under
the assumption $\sin\theta_{13}=0$.
Here $M_2$ and $|h_2|$ are fixed to typical values such as
$M_2=1.01M_\eta$ and $|h_2|=10^{-3.5}$. 
As long as $|h_1|\gg |h_2|$ is satisfied, we find that 
the value of $|h_1|$ is not varied by changing the value of $|h_2|$.
This figure shows that larger values of $M_\eta$ require 
larger values of neutrino Yukawa couplings $|h_{1,3}|$ 
to satisfy the conditions in eq.~(\ref{c-oscil}) when $M_{1,3}$ 
and $\lambda_5$ are fixed.
We note that neutrino Yukawa couplings $|h_{1,3}|$ can take values 
such as $O(10^{-3})$ when we fix $|\lambda_5|$ in suitable ranges.
This is favorable for the thermal leptogenesis as seen below.
 
\input epsf
\begin{figure}[t]
\begin{center}
\epsfxsize=7cm
\leavevmode
\epsfbox{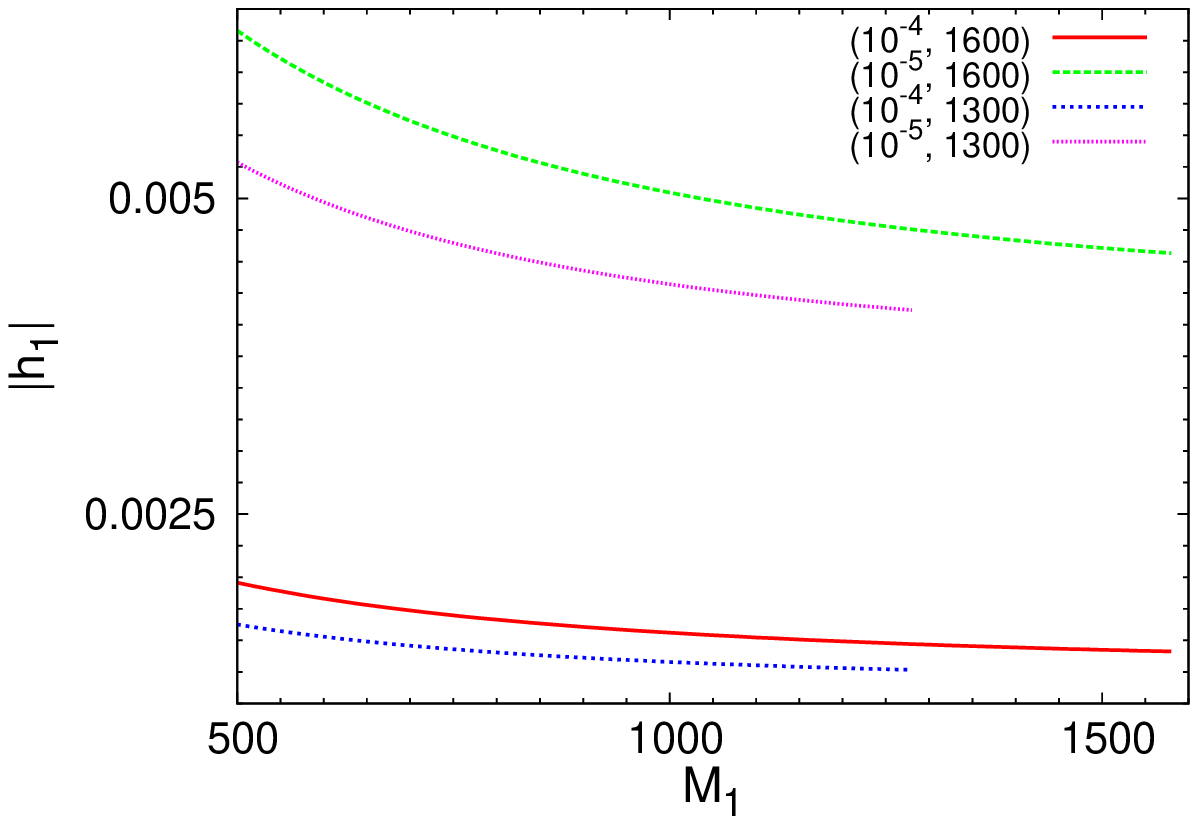}
\hspace*{5mm}
\epsfxsize=7cm
\leavevmode
\epsfbox{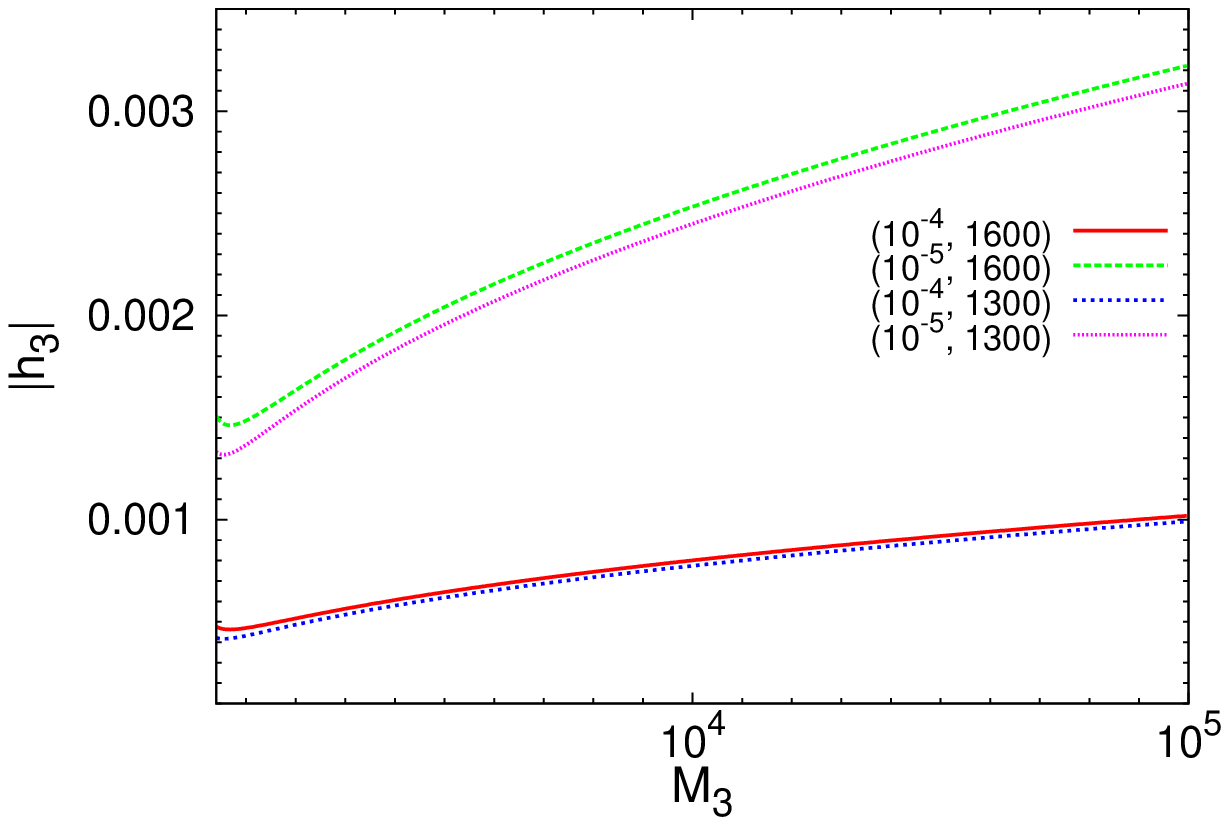}
\end{center}
\vspace*{-3mm}

{\footnotesize {\bf Fig.~1}~~Neutrino Yukawa couplings $|h_{1,3}|$
 which satisfy the neutrino oscillation data. Each line is
 plotted as a function of $M_{1,3}$ for typical values of 
$(\lambda_5, M_\eta)$ which are shown in the figures.
A GeV unit is used for the mass scale.}  
\end{figure}

Now we consider leptogenesis in this model.
If we suppose a situation such that $\eta$ has no lepton number and 
all the right-handed neutrinos decouple at some TeV region, 
$B-L$ is conserved below this decoupling temperature 
as in the canonical leptogenesis.\footnote{
One may consider the lepton number $L$ defined as $L(\eta)=1$ and
$L(N_{R_i})=0$. In this case the lepton number is violated by the
$\lambda_5$ term in eq.~(\ref{model}). Since $\lambda_5$ should be
small enough for this term to decouple at $T>100$~GeV \cite{deflm}, 
it seems to be difficult to cause sufficient CP asymmetry 
in the decay of thermal $N_{R_2}$. Thus, we do not consider it here.}
Thus, if the excess of the number density of $N_{R_i}$ over the 
equilibrium value is caused before the freeze-out of the 
sphaleron interaction, the baryon number asymmetry $n_B$ is expected to 
be processed from the lepton number asymmetry 
$n_L(\equiv n_\ell-n_{\bar\ell})$ generated through the $N_{R_i}$ decay. 
If we represent the ratio of baryon number asymmetry $n_B$ 
to an entropy density $s(\equiv\frac{2\pi^2}{45}g_\ast T^3)$ 
as $Y_B$, it is calculated by using the 
lepton number asymmetry $Y_L(\equiv \frac{n_L}{s})$ as
\begin{equation}
Y_B=-\frac{8}{23}Y_L(z_{\rm EW}),
\label{baryon}
\end{equation}
where $z_{\rm EW}$ is related to the sphaleron decoupling temperature 
$T_{\rm EW}$ through $z_{\rm EW}=\frac{M_2}{T_{\rm EW}}$. 
In eq.~(\ref{baryon}), we use $B= \frac{8}{23}(B-L)$ which
is satisfied also in this model.

Since $N_{R_1}$ is stable due to the $Z_2$ symmetry, leptogenesis 
based on the decay of the thermal $N_{R_1}$ is not allowed.
Thus, the lepton number asymmetry is expected to be 
produced through the decay of $N_{R_2}$ whose dominant mode is 
$N_{R_2}\rightarrow \ell_\alpha\eta^\dagger$. 
One might expect the enhancement of CP asymmetry in this decay 
due to the degenerate mass spectrum (\ref{nspec}) as in the ordinary 
resonant leptogenesis \cite{resonant,resonant1}. However, 
we should note that it can not occur here since $N_{R_1}$ is stable. 
In this model there are also several dangerous lepton number violating 
processes which wash out the generated lepton number asymmetry.
In the Appendix we present the formulas of the reaction 
density $\gamma$ relevant to the calculation of $Y_L$.  
The Boltzmann equations for 
$Y_{N_{R_2}}\left(\equiv\frac{n_{N_{R_2}}}{s}\right)$ and $Y_L$ are 
written as \cite{kt,cross}\footnote{Since $N_{R_{1,3}}$ and $\eta$ 
have sufficient interactions due to the Yukawa interactions with the 
couplings $h_{1,3}$ and the weak interaction, they are considered to be 
in the thermal equilibrium during these evolution. }
\begin{eqnarray}
\frac{dY_{N_{R_2}}}{dz}&=&-\frac{z}{sH(M_2)}
\left(\frac{Y_{N_{R_2}}}{Y_{N_{R_2}}^{\rm eq}}-1\right)\left[\gamma_D^{N_2}
+\sum_{i=1,3}\left(\gamma_{{N_2}{N_i}}^{(2)}
+\gamma_{{N_2}{N_i}}^{(3)}\right)\right], \nonumber \\
\frac{dY_L}{dz}&=&\frac{z}{sH(M_2)}\left[
\varepsilon\left(\frac{Y_{N_{R_2}}}{Y_{N_{R_2}}^{\rm
	    eq}}-1\right)\gamma_D^{N_2}
-\frac{2Y_L}{Y_\ell^{\rm eq}}\left(\frac{\gamma_D^{\eta}}{4}+\gamma_N^{(2)}
+\gamma_N^{(13)}\right)\right], 
\label{bqn}
\end{eqnarray}
where $H(M_2)=1.66g_\ast^{-1/2}\frac{M_2^2}{m_{\rm pl}}$, 
$Y_\ell^{\rm eq}=\frac{45}{\pi^4g_\ast}$ and $Y_{N_{R_2}}^{\rm eq}$
stands for the equilibrium value of $Y_{N_{R_2}}$. 
In these Boltzmann equations we omit several terms whose contributions 
are negligible compared with others. 

The CP asymmetry $\varepsilon$ induced in the $N_{R_2}$ decay 
comes from the interference between a tree diagram and 
one-loop vertex or self-energy diagrams as is well known.\footnote{
It is useful to note that the lepton number asymmetry might 
be considered to be generated through the decay of $\eta$ also. 
However, since $\eta$ is kept in the
thermal equilibrium through various interactions, its out of equilibrium
decay does not occur and the lepton number asymmetry is not expected to
be produced through its decay.}
It can be expressed for the mass spectrum (\ref{nspec}) as \cite{resonant1}
\begin{eqnarray}
\varepsilon&=&\frac{1}{16\pi
\left[\frac{3}{4}+\frac{1}{4}\left(1-\frac{M_\eta^2}{M_2^2}\right)^2\right]}
\sum_{i=1,3}\frac{\displaystyle {\rm Im}
\left[\left(\sum_{k=e,\mu,\tau}h_{k2}h^\ast_{ki}\right)^2\right]}
{\displaystyle \sum_{k=e,\mu,\tau}h_{k2}h_{k2}^\ast}
G\left(\frac{M_i^2}{M_2^2},\frac{M_\eta^2}{M_2^2}\right)\nonumber \\
&=&\frac{1}{16\pi\left[\frac{3}{4}+\frac{1}{4}
\left(1-\frac{M_\eta^2}{M_2^2}\right)^2\right]}
\left[\frac{4}{2+C^2}|h_1|^2
G\left(\frac{M_1^2}{M_2^2},\frac{M_\eta^2}{M_2^2}\right)
\sin 2(\varphi_2-\varphi_1)\right. \nonumber \\
&&\hspace*{4cm}\left.+
\frac{C^2}{2+C^2}|h_3|^2
G\left(\frac{M_3^2}{M_2^2},\frac{M_\eta^2}{M_2^2}\right)
\sin 2(\varphi_2-\varphi_3)\right] \nonumber \\
&\equiv&\varepsilon_1\sin 2(\varphi_2-\varphi_1)+
\varepsilon_3\sin 2(\varphi_2-\varphi_3).
\label{cp}
\end{eqnarray}
In this formula $G(x,y)$ is expressed as
\begin{equation}
G(x,y)=\frac{5}{4}F(x,0)+\frac{1}{4}F(x,y)
+\frac{1}{4}(1-y)^2\left[F(x,0)+F(x,y)\right], 
\end{equation}
where $F(x,y)$ is defined by
\begin{equation}
F(x,y)=\sqrt{x}\left[1-y-(1+x)\ln\left(\frac{1-y+x}{x}\right)\right].
\end{equation}
The flavor structure (\ref{yukawa}) is used in this derivation.
The magnitude of $\varepsilon$ is determined by the values of
$\varepsilon_{1,3}$ and $\sin 2(\varphi_2-\varphi_{1,3})$.
Since the CP asymmetry induced in the decay of $N_{R_2}$ is required to
have sufficient magnitude, $|h_{1,3}|$ should not be largely suppressed. 
Here we should remind that $|h_{1,3}|$ can take wide range values to 
generate the appropriate neutrino masses by varying a value of
$\lambda_5$ as shown in Fig.~1.
Because of this feature of the radiative neutrino mass model, the
neutrino Yukawa couplings can take appropriate values for the CP
asymmetry even for the TeV scale right-handed neutrinos. 
Numerical values of $|h_{1,3}|$ and $\varepsilon_{1,3}$ 
are presented in Table 1 for typical values of model parameters. 

\begin{figure}[t]
\begin{center}
\begin{tabular}{|c|c|c|c|c|c|c|c|c|c|}\hline
&$\lambda_5$& $M_\eta$ & $C$  & $10^{-3}|h_1|$ & $10^{-3}|h_3|$ 
&$10^7\varepsilon_1$ &$10^{10}\varepsilon_3$ & $10^{10}|Y_B|$ & 
$|\sin\theta_{13}|$ \\ \hline
(a)&$10^{-4}$ &1.35 &$-0.5$ &1.23 &1.12 &$-0.379$ & $-0.229$  & 0.28 & 
0.087    \\ \hline
(b)&$10^{-4}$ &1.60& $-0.5$ &1.41 &1.21 & $-0.453$ & $-0.272$ & 0.32 &
0.074  \\ \hline
(c)&$10^{-4}$ &1.80 &$-0.5$ &1.50 &1.29  & $-0.513$ & $-0.306$ & 0.34 &
0.065   \\ \hline
(d)&$10^{-5}$ &1.60 &$-1$ &4.56 &3.84  & $-3.55$ & $-8.16$ & 0.30  &
0.015  \\ \hline
\end{tabular}
\end{center}
\vspace*{2mm}
{\footnotesize{\bf Table 1}~~The CP asymmetry parameters 
$\varepsilon_{1,3}$ and baryon number asymmetry $|Y_B|$ predicted for typical 
parameters which satisfy the constraints from the neutrino oscillation data.  
Remaining parameters are fixed as $|h_2|=10^{-3.5}$, $M_3=100M_\eta$,
$\Delta_1=10^{-5}$ and $\Delta_2=10^{-3}$, where 
$\Delta_i$ is defined by $\Delta_i=\frac{|M_\eta-M_i|}{M_\eta}$. 
In the estimation of $|Y_B|$, the maximum value of 
$|\sin 2(\varphi_2-\varphi_1)|$ is assumed. A TeV unit is used for 
the mass scale.}
\end{figure} 

It is useful to discuss nature of the favorable parameters before 
presenting the numerical results of the Boltzmann equations. 
In order to generate the lepton number asymmetry effectively, 
a sufficient number of $N_{R_2}$ should be successfully produced as 
the out of thermal equilibrium states. 
The lepton number violating processes induced by the light 
right-handed neutrinos should also be sufficiently suppressed 
keeping the CP asymmetry $\varepsilon$ the appropriate value.
We have to choose parameters which satisfy these simultaneously.
As such a promising situation, we consider the one where the inverse decay 
of $\eta$ could be the dominant mode for the lepton number violating 
processes. Such a case is expected to occur for finely degenerate 
$M_1$, $M_\eta$ and $M_2$. In the following analysis we assume 
the degenerate masses such as $\Delta_1=10^{-5}$ and $\Delta_2=10^{-3}$, 
where $\Delta_i$ is defined by $\Delta_i=\frac{|M_\eta-M_i|}{M_\eta}$. 

In order to confirm this, we study the behavior of the ratio of
thermally averaged reaction rate $\langle\Gamma(z)\rangle$ to
Hubble parameter $H(z)$ for each relevant processes by assuming the
above mentioned degenerate mass spectrum. 
The thermally averaged reaction rate 
$\langle\Gamma\rangle$ corresponding to the reaction 
density $\gamma$ included in eq.~(\ref{bqn}) are given as
\begin{equation}
\langle\Gamma_D^{N_2}\rangle=\frac{\gamma_D^{N_2}}
{n_{N_{R_2}}^{\rm eq}}, 
\qquad
\langle\Gamma_{ID}^\eta\rangle=\frac{\gamma_D^{\eta}}{n_\ell^{\rm eq}} 
\end{equation}
for the decay of $N_{R_2}$ and the inverse decay of $\eta$, and also
\begin{equation}
\langle\Gamma_N^{(2,13)}\rangle=
\frac{\gamma_N^{(2,13)}}{n_\ell^{\rm eq}}, \qquad  
\langle\Gamma_{N_iN_2}^{(2,3)}\rangle=
\frac{\gamma_{N_iN_2}^{(2,3)}}{n_{N_{R_2}}^{\rm eq}}
\end{equation}
for the 2-2 scattering processes given in eqs.~(\ref{lv1}), (\ref{lv2})
and eqs.~(\ref{nlv1}), (\ref{nlv2}), respectively.
If both $\frac{\langle\Gamma_{ID}^{\eta}\rangle}{H(M_2)}$ and 
$\frac{\langle\Gamma_N^{(2,13)}\rangle}{H(M_2)}$ are of $O(1)$ at
a neighborhood of $z\sim 1$, we find that the favorable situation for the 
leptogenesis could be realized for the assumed degenerate masses. 
In Fig.~2, we plot $\frac{\langle\Gamma\rangle}{H}$ as a function of $z$ 
for each process. 
This figure shows that all the lepton number violating processes 
could be out of thermal equilibrium at the period $1 <z < 6$.
Although these dangerous processes are mediated by rather light
right-handed fields, we find that the mass degeneracy could play a 
crucial role to make them ineffective due to the Boltzmann suppression. 
It could compensate the disfavored effect brought by the lightness 
of the right-handed neutrinos.
It should be noted that the suppression of the washout effect of 
the lepton number asymmetry in this model is brought as the 
combined effect of the degenerate masses and the neutrino Yukawa couplings 
of $O(10^{-3})$.  
 
\begin{figure}[t]
\begin{center}
\epsfxsize=7cm
\leavevmode
\epsfbox{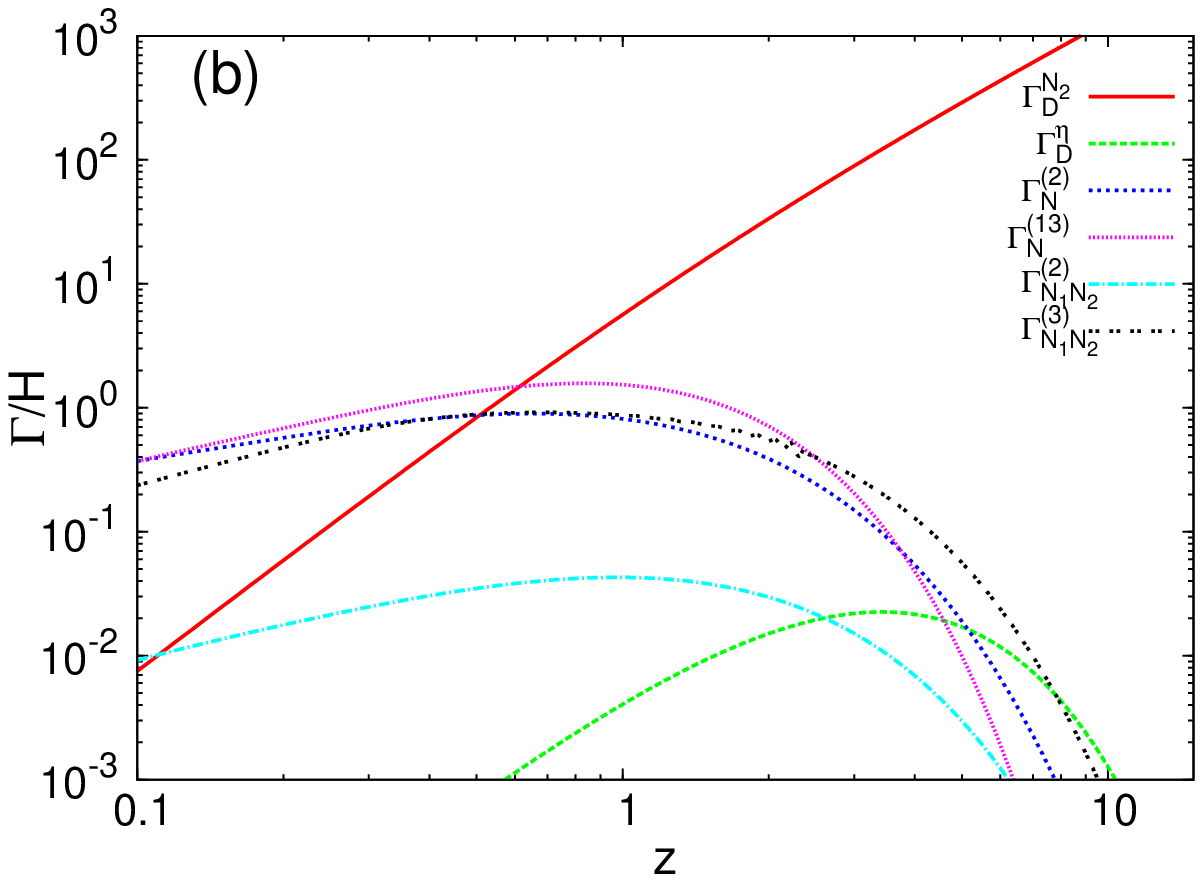}
\hspace*{5mm}
\epsfxsize=7cm
\leavevmode
\epsfbox{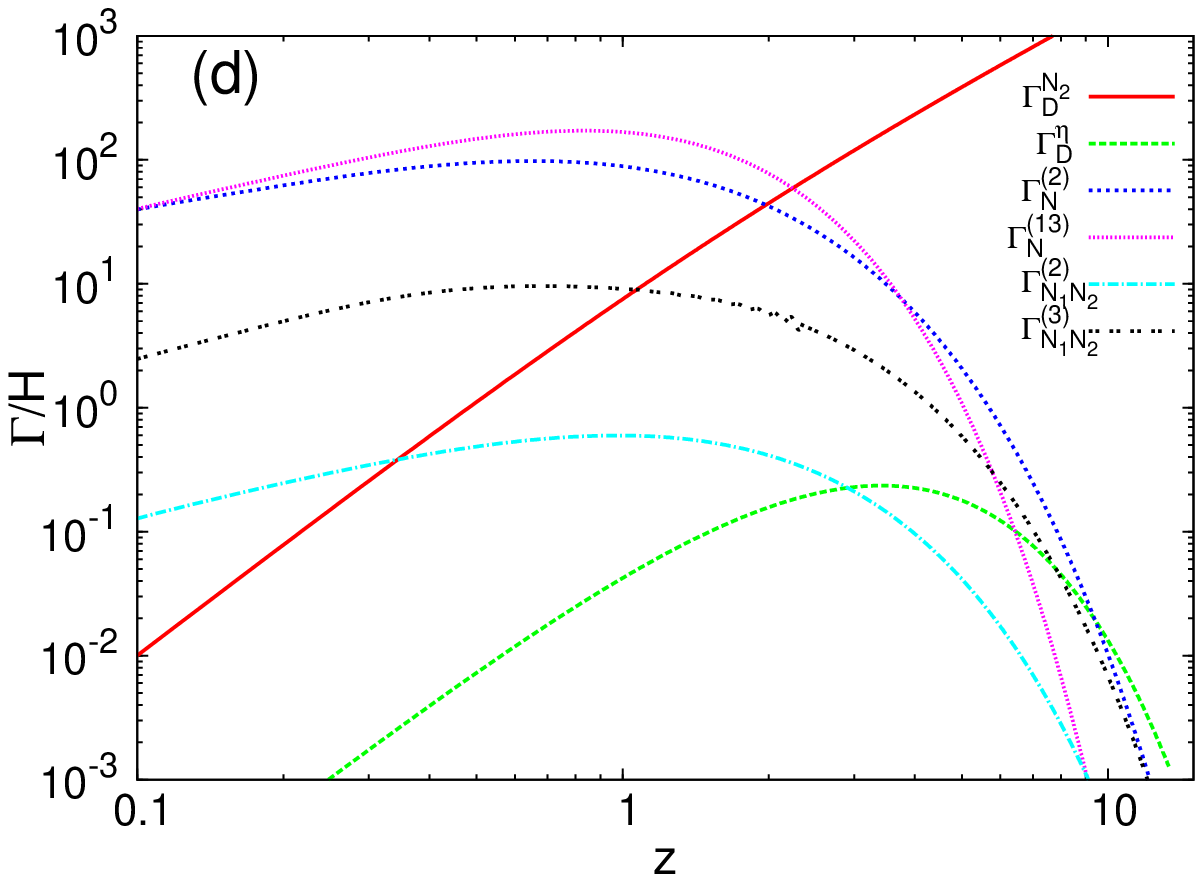}
\end{center}
\vspace*{-3mm}

{\footnotesize {\bf Fig.~2}~~ Relevant reaction rates as functions of
 $z$. In each figure the model parameters are fixed to the ones in 
the cases (b) and (d) in Table 1.}  
\end{figure}

Here we give our numerical results for the generated baryon number
asymmetry in the present model.
In Fig.~3, we show the evolution of $Y_{N_{R_2}}$,
$\Delta_{N_{R_2}}(\equiv|Y_{N_{R_2}}-Y_{N_{R_2}}^{\rm eq}|)$ 
and $|Y_L|$ which are obtained by solving the Boltzmann 
equations (\ref{bqn}) for the cases (b) and (d) listed in Table~1.
In this calculation we assume that $\sin 2(\varphi_2-\varphi_1)$ 
takes a maximum value, for simplicity.    
The figure shows that $|Y_L|$ reaches a constant value before the sphaleron
decoupling if we consider the sphaleron decoupling temperature as 
$T_{\rm EW}\sim 140$~GeV. This temperature $T_{\rm EW}$ corresponds 
to the Higgs mass such as 125 GeV \cite{sphtem}, which could be 
considered as the promising one from the recent ATLAS and CMS data. 
The obtained $Y_B$ through this analysis is also listed in Table~1 for
each case. Their order is correct but the values are a little bit 
smaller than the one expected from the big bang nucleosynthesis. 
However, we should remind that they are
obtained for the very simple flavor structure of neutrino Yukawa couplings.
The model could also give nonzero values for $\sin\theta_{13}$ 
if $C\not=0$ is assumed. The predicted value of $|\sin\theta_{13}|$ 
for each case is also listed in Table 1.
These results could be changed by modifying the flavor structure of
neutrino Yukawa couplings.
If we adopt other type of flavor structure which can be consistent 
with the neutrino oscillation data, desired values for both $Y_B$ and
$|\sin\theta_{13}$ might be derived within this framework.  
Although this study is an interesting subject, it is beyond the scope of
this paper. 

\begin{figure}[t]
\begin{center}
\epsfxsize=7.5cm
\leavevmode
\epsfbox{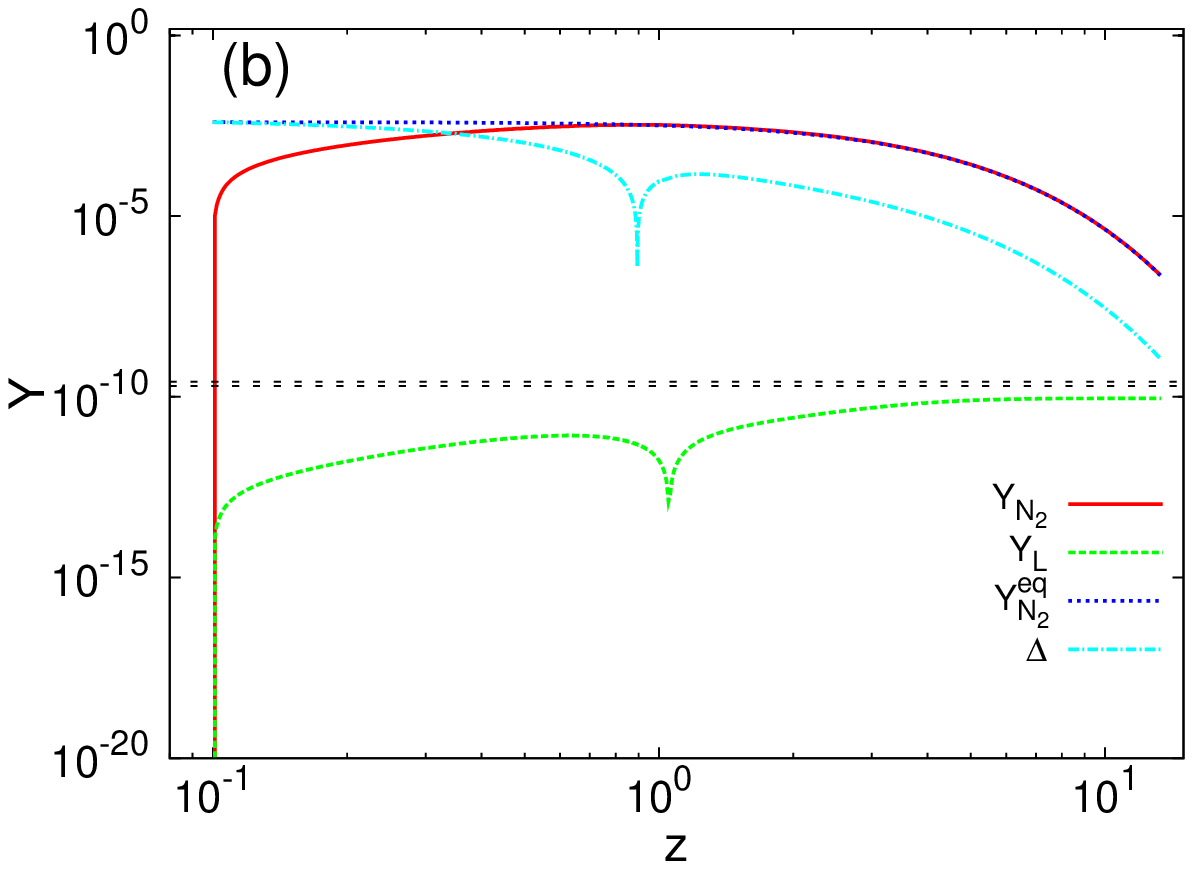}
\hspace*{1mm}
\epsfxsize=7.5cm
\leavevmode
\epsfbox{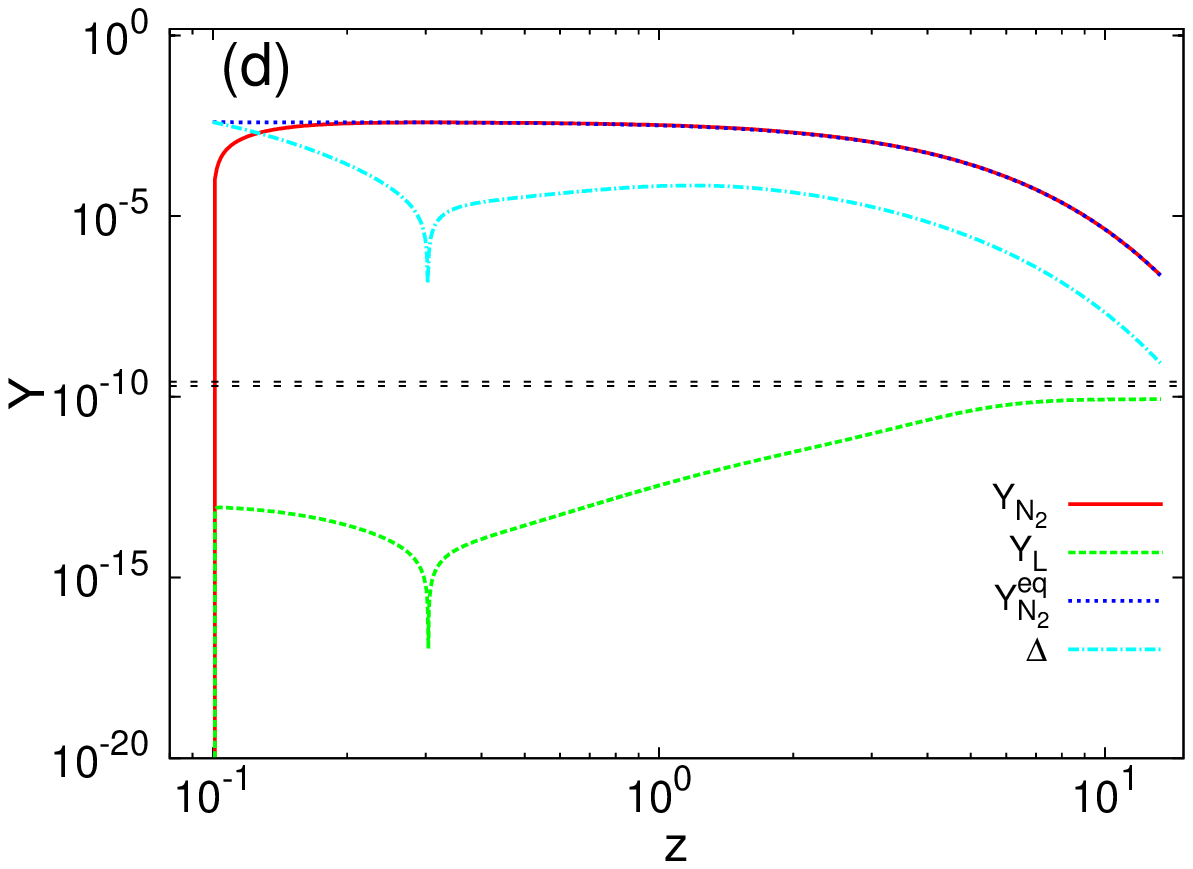}
\end{center}
\vspace*{-3mm}

{\footnotesize {\bf Fig.~3}~~The evolution of $Y_{N_{R_2}}$ and 
$|Y_L|$ generated through the $N_{R_2}$ decay. Each figure is plotted
 for the cases (b) and (d) given in Table 1. 
Black dashed line represents the value of $|Y_L|$ required to 
realize the observed baryon number asymmetry $Y_B=(0.7-0.9)\times 10^{-10}$.}  
\end{figure}

This leptogenesis scenario is characterized by the requirement that 
the masses of $N_{R_{1,2}}$ and $\eta$ should be finely degenerate.
This situation may be considered similar to the resonant leptogenesis.
However, the required mass degeneracy is much milder and 
also has different nature compared with the one of the resonant 
leptogenesis where the required mass degeneracy among the right-handed
neutrinos is smaller than $10^{-10}$ \cite{resonant}.
We note that this difference is caused by the neutrino mass generation
mechanism, especially, the existence of the small coupling $\lambda_5$. 
Since the same Yukawa couplings contribute to both the 
generation of the lepton number asymmetry and its washout, 
some extra suppression for the washout processes is required 
to keep the lepton number asymmetry in the favorable range. 
In the present model, this effect could be brought through 
the mass degeneracy of the relevant fields which suppress both 
the relevant decay and also the lepton number violating 
scattering processes.
As long as the reheating temperature satisfies $T>M_2$, the $N_{R_2}$
could be in the thermal equilibrium. 
Thus, we find that rather low reheating temperature
such as $10^{5}$ GeV could make this leptogenesis scenario applicable.  

\section{Phenomenological constrains}
We have seen that the suitable lepton number asymmetry could 
be thermally generated in this radiative neutrino mass model
for a rather low reheating temperature. 
In this section we study this possibility further 
by taking account of other phenomenological constraints 
in a quantitative way.

First, we start to address a problem relevant to the CP phase,
which eventually appears when we consider the leptogenesis. 
In general, the interactions introduced to generate the neutrino masses 
could also contribute to the electric dipole moment of an electron (EDME) 
through one-loop diagrams if they violate the CP invariance. 
However, as long as we confine the additional interaction for leptons 
to the one given in eq.~(\ref{model}), we find that the EDME is 
not induced at one-loop level but can be induced through a two-loop 
diagram with internal $W^\pm$ lines. Even if the Majorana phases
$\alpha_{1,2} $ given in eq.~(\ref{cpphase}) are assumed to take 
a maximum value, the EDME ($d_e/e$) induced by such a diagram is roughly
estimated to be $O(G_F^2m_e^3)$. It is much smaller than the present
experimental upper bound \cite{edm}. Thus, the constraint from the EDME
does not contradict with the present leptogenesis scenario.

Second, the lepton flavor violating processes such as
$\ell_i\rightarrow \ell_j\gamma$ are induced through one loop
diagrams similar to the ones for the neutrino masses. However, 
since they are irrelevant to $\lambda_5$ unlike the neutrino masses, 
large contributions could be generated from them. 
In fact, the relevant branching ratio can be expressed as \cite{raddm2}
\begin{eqnarray}
&&{\rm Br}(\mu\rightarrow e\gamma)\simeq \frac{3\alpha}{64\pi(G_FM_\eta^2)^2}
\left[C|h_2|^2F_2\left(\frac{M_2^2}{M_\eta^2}\right)
+|h_3|^2F_2\left(\frac{M_3^2}{M_\eta^2}\right)\right]^2, \nonumber \\
&&{\rm Br}(\tau\rightarrow\mu\gamma)\simeq 
\frac{0.51\alpha}{64\pi(G_FM_\eta^2)^2}
\left[\left(|h_1|^2+|h_2|^2\right)F_2\left(\frac{M_1^2}{M_\eta^2}\right)
-|h_3|^2F_2\left(\frac{M_3^2}{M_\eta^2}\right)\right]^2, 
\label{blfv}
\end{eqnarray}
where $F_2(r)$ is defined as
\begin{equation}
F_2(r)=\frac{1-6r+3r^2+2r^3-6r^2\ln r}{6(1-r)^4}.
\end{equation} 
In this derivation we use $M_1\sim M_2$.
It is known that these processes severely constrain the model if the
neutrino Yukawa couplings take values of $O(1)$ which are  convenient
for the explanation of the relic abundance of the lightest right-handed
neutrino.
However, the neutrino Yukawa couplings have rather small values such as
$O(10^{-3})$ in the cases of Table~1, eq.~(\ref{blfv}) gives negligibly
small values compared with the present experimental upper bounds \cite{lfv}.
 
Finally, we examine the consistency of the scenario with 
the DM relic abundance.
Since $N_{R_1}$ is the lightest $Z_2$ odd particle, it is stable to be
DM. Thus, its relic abundance should satisfy 
$\Omega_{N_{R_1}}h^2=0.11$ which is
obtained from observations of the WMAP \cite{wmap}. 
Here we remind that $N_{R_1}$ interacts with other fields only through
the neutrino Yukawa coupling $h_1$.
Although it is required to have values of $O(1)$ to realize the favored
value of $\Omega_{N_{R_1}}h^2$ as shown in the 
previous work \cite{raddm1,raddm2}, the above study suggests that 
it should be much smaller to explain both the neutrino oscillation 
data and the baryon number asymmetry in the universe. 
The relic abundance of $N_{R_1}$ is so large to overclose the universe
in the present scenario.\footnote{Since $M_1$ are degenerate with $M_2$ and
$M_\eta$, the coannihilation of $N_{R_1}$ with $N_{R_2}$ and $\eta$
should be taken into account. However, even in that case 
the neutrino Yukawa couplings $h_{1,2}$ are too small to reduce 
its relic abundance sufficiently.}
 
This problem might be solved by introducing some interaction which 
makes $N_{R_1}$ unstable but induces no other substantial
effect. As such a simple example, we may consider the gravity induced 
$Z_2$ violating interaction similar to the Weinberg operator
\cite{weinberg} such as
\begin{equation}
{\cal L}_V=\frac{f_i}{M_{\rm pl}}(\ell^T_i\phi^\ast)(\eta^\dagger\ell_i) 
+{\rm h.c.}.
\end{equation} 
This interaction brings three body decay
$N_{R_1}\rightarrow\bar\ell\ell\ell$ for $N_{R_1}$.
In this case the lifetime of $N_{R_1}$ is estimated as
\begin{equation}
\tau_{N_{R_1}}\simeq 3\times 10^{16}\left(\frac{1~{\rm TeV}}{M_1}\right)
\left(\frac{10^{-3}}{|h_1|}\right)^2 ~{\rm sec}
\end{equation}
for $f_i=O(1)$.
Since this lifetime is shorter than the age of universe,
$N_{R_1}$ can not be DM and then we need to introduce a new DM candidate.   
If we consider the embedding of the thermal leptogenesis in this 
radiative neutrino mass model in the simple way,
the model seems to lose the close relation between the neutrino masses and 
the existence of DM generally.
We order several comments on this point.
 
As long as we confine our consideration to the thermal leptogenesis in
this radiative neutrino mass model, we can only explain two 
of three experimental results which suggest physics beyond the SM, 
that is, the neutrino oscillation data 
and the DM relic abundance as discussed in \cite{raddm1,raddm2,raddm3}, 
or the neutrino oscillation data and the baryon number asymmetry 
as studied here.
In the former case, we need to find some new generation mechanism of 
the baryon number asymmetry. The promising scenario is nonthermal
leptogenesis, which has been discussed in \cite{ad,deflm}.

In the latter case, we need to modify the model so as to include
some new scenario for DM. We discuss two scenarios here.
The first one is to introduce additional interaction which
contribute to the pair annihilation of $N_{R_1}$.
As such an example, one might consider an flavor blind abelian gauge 
interaction at TeV regions, which induces $s$-channel pair annihilation 
of $N_{R_1}$. This type of model has been discussed in \cite{raddm2}.
Unfortunately, this extension seems to make the thermal leptogenesis 
useless, since the same interaction also contributes the pair
annihilation of $N_{R_2}$ which keeps $N_{R_2}$ in the thermal
equilibrium until larger $z$.
Another promising solution for this issue is to introduce 
an interaction $y_iS\bar N_{R_i}^cN_{R_i}$ with 
a $Z_2$ even scalar field $S$, which has been studied in other 
context in \cite{raddm3}.
If Yukawa couplings satisfy $|y_1|\gg |y_{2,3}|$, this could mainly 
contribute to the pair annihilation of $N_{R_1}$ through $s$-channel
exchange of $S$. It could reduce the relic abundance of $N_{R_1}$ 
substantially due to the resonance as long as the mass of $S$ satisfies 
the condition $m_S^2\simeq 4M_1^2$.
Detailed study of this issue will be given elsewhere.
 
The second one is to construct a supersymmetric version of the 
present model. It might be done straightforwardly 
following the prescription shown in \cite{susyrad}.
Thermal leptogenesis is expected to be formulated 
along the similar line to the present model. 
An interesting point in this extension is that the artificial 
hierarchy assumed for various couplings and masses might be derived on the 
basis of symmetry. In fact, if we introduce an anomalous U(1) 
symmetry at high energy regions, the required hierarchical 
structure of the couplings and masses might also be generated 
via its spontaneous breaking as discussed in \cite{susyrad}.
This extension may also open a new possibility for DM. Since $N_{R_1}$ can be
unstable there, the lightest neutralino could be dominant component of DM. 
In this extended model, we might have a supersymmetric model with 
no gravitino problem since the required reheating temperature could be 
$10^{5}$ GeV as shown here. 
These extensions of the model might give a simultaneous explanation 
within the thermal leptogenesis framework
for the three crucial problems in the SM, that is, the baryon number 
asymmetry in the universe, the small neutrino masses consistent with the
neutrino oscillation data, and the DM relic abundance. 

\section{Summary}
The radiative neutrino mass model has been originally proposed as 
the model which could give the consistent explanation for the neutrino 
masses and mixings and also the relic abundance of DM on the basis 
of TeV scale physics.
If we assume the simple flavor structure for the neutrino Yukawa 
couplings in this model, the tri-bimaximal neutrino mixing and
the suppression of lepton flavor violating processes such as 
$\mu\rightarrow e\gamma$ can be easily derived. 
However, unfortunately, it is not so easy to embed the thermal 
leptogenesis in this model because of the lightness of 
right-handed neutrinos.\footnote{ 
We should remind that the original Ma model can be consistent with 
the thermal leptogenesis if the the lightest neutral component 
of the inert doublet $\eta$ is identified as DM \cite{ma1}. However, the 
right-handed neutrino should be of $(10^{7})$ GeV or more in that scenario.}  
In this paper we have studied under what condition
thermal leptogenesis could be applicable in the model with the
right-handed neutrinos of $O(1)$ TeV mass 
for the explanation of the baryon number asymmetry in the universe. 
Our result is that the finely degenerate mass spectrum of $Z_2$
odd fields could allow the model to generate the sufficient baryon 
number asymmetry by thermal leptogenesis, although the masses of
right-handed neutrinos are not huge but of $O(1)$ TeV.
Since the relevant right-handed neutrino is light, the reheating 
temperature could be less than $10^5$ GeV. 
This suggests that the supersymmetric extension of the model could 
escape the notorious gravitino problem. 

As shown in this paper, unfortunately, the close relation between 
the neutrino masses and the existence of DM is lost if we try to 
embed the thermal leptogenesis in
the model. The neutrino Yukawa couplings required by the thermal
leptogenesis are too small to reduce the relic abundance of the 
DM candidate $N_{R_1}$.
We need some extension of the model to resolve this problem by
introducing suitable interaction for $N_{R_1}$ which makes $N_{R_1}$
unstable or contributes to the $N_{R_1}$ annihilation.  
Anyway, an interesting point is that the extension discussed in this
paper could give a closely correlated explanation for recently clarified 
phenomena relevant to physics beyond the SM.  
They seem to deserve further study. 

\section*{Acknowledgement}
This work is partially supported by a Grant-in-Aid for Scientific
Research (C) from Japan Society for Promotion of Science (No.21540262)
and also a Grant-in-Aid for Scientific Research on Priority Areas 
from The Ministry of Education, Culture, Sports, Science and Technology 
(No.22011003).

\newpage
\section*{Appendix}
In this appendix we give the formulas of the reaction density of the
relevant processes. Since $N_{R_{1,3}}$ and $\eta$ are considered to be
in the thermal equilibrium due to the Yukawa interactions
and other interactions, we need to study the Boltzmann equations only 
for the number density of $N_{R_2}$ and the lepton number asymmetry 
as discussed in the text.  
For the processes relevant to their evolution, we can refer to 
the reaction density given in \cite{cross}.
However, interaction terms of $\eta$ and $N_{R_2}$ are restricted by the
$Z_2$ symmetry as shown in eq.(\ref{model}). It causes large 
difference from the ordinary seesaw leptogenesis.  
We need to modify them by taking account of the features 
of the present model such that $\eta$ has a large mass 
comparable with the one of $N_{R_{1,2}}$ and the neutrino Yukawa 
couplings have the flavor structure given in eq.~(\ref{yukawa}).
In order to give the expression for the reaction density of the relevant
processes, we introduce dimensionless variables
\begin{equation}
x=\frac{s}{M_2^2}, \qquad a_j=\frac{M_j^2}{M_2^2}, \qquad 
a_\eta=\frac{M_\eta^2}{M_2^2}.
\end{equation}
where $s$ is the squared center of mass energy.

The reaction density for the decay of $N_{R_2}$ and $\eta$ can be 
expressed as
\begin{eqnarray}
&&\gamma_D^{N_2}=\frac{|h_2|^2(2+C^2)}{8\pi^3}M_2^4
\left(1-a_\eta\right)^2\frac{K_1(z)}{z}, \nonumber \\ 
&&\gamma_D^{\eta}=\frac{|h_1|^2}{8\pi^3}a_\eta^{3/2}M_\eta^4
\left(1-\frac{a_1}{a_\eta}\right)^2
\frac{K_1\left(\sqrt{a_\eta}z\right)}{z}, 
\label{decay}
\end{eqnarray} 
where $K_1(z)$ is the modified Bessel function of the second kind.
The reaction density for the scattering processes are expressed as
\begin{equation}
\gamma(ab\rightarrow ij)=\frac{T}{64\pi^4}\int^\infty_{s_{\rm min}}ds~
\hat\sigma(s)\sqrt{s}K_1\left(\frac{\sqrt{s}}{T}\right),
\end{equation}
where $s_{\rm min}={\rm max}[(m_a+m_b)^2,(m_i+m_j)^2]$ and 
$\hat\sigma(s)$ is the reduced cross section. 
In order to express the reduced cross section for the scattering 
processes relevant to eq.~(\ref{bqn}), we define the quantities as
\begin{eqnarray}
&&\frac{1}{D_1(x)}=\frac{1}{x-a_1}, \qquad
\frac{1}{D_3(x)}=\frac{x-a_3}{(x-a_3)^2+a_3c_3}, \qquad 
c_3=\frac{9a_3}{16\pi^2}|h_3|^4, \nonumber \\
&&\lambda_{ij}=\left[x-(\sqrt{a_i}+\sqrt{a_j})^2\right]
\left[x-(\sqrt{a_i}-\sqrt{a_j})^2\right],
\nonumber \\
&&L_{ij}=\ln\left[\frac{x-a_i-a_j+ 2a_\eta +\sqrt{\lambda_{ij}}}
{x-a_i-a_j +2 a_\eta -\sqrt{\lambda_{ij}}}\right], \nonumber \\
&&L_{ij}^\prime=\ln\left[\frac{\sqrt{x}(x-a_i-a_j-2a_\eta)
+\sqrt{\lambda_{ij}(x-4a_\eta)}}
{\sqrt{x}(x-a_i-a_j-2a_\eta) 
-\sqrt{\lambda_{ij}(x-4a_\eta)}}\right].
\end{eqnarray}
The relevant reduced cross section are summarized as follows.

As the lepton number violating scattering processes induced through the
$N_{R_{1,3}}$ exchange,
we have
\begin{eqnarray}
\hat\sigma^{(2)}_N(x)&=&\frac{1}{2\pi}\left[4|h_1|^4
\frac{(x-a_\eta)^2a_3}{x^3}
\left\{\frac{x^2}{xa_1 -a_\eta^2}+\frac{2x}{D_1(x)}
+\frac{(x-a_\eta)^2}{2D_1(x)^2}\right.\right.\nonumber \\
&-&\left.\frac{x^2}{(x-a_\eta)^2}
\left(1+\frac{2(x+a_1)-4a_\eta}{D_1(x)}\right)
\ln\left(\frac{x(x+a_1-2a_\eta)}{xa_1-a_\eta^2}\right)\right\}\nonumber \\
&+&\left.
9|h_3|^4\frac{(x-a_\eta)^2a_3}{x^3}\left\{
\frac{x^2}{xa_3 -a_\eta^2}+\frac{2x}{D_3(x)}+\frac{(x-a_\eta)^2}{2D_3(x)^2}
\right.\right.\nonumber \\
&-&\left.\left.\frac{x^2}{(x-a_\eta)^2}
\left(1+\frac{2(x+a_3)-4a_\eta}{D_3(x)}\right)
\ln\left(\frac{x(x+a_3-2a_\eta)}{xa_3-a_\eta^2}\right)\right\}
\right] 
\label{lv1}
\end{eqnarray}
for $\ell_\alpha\eta^\dagger \rightarrow \bar\ell_\beta\eta$ and also
\begin{eqnarray}
\hat\sigma^{(13)}_N(x)&=&\frac{1}{2\pi}\frac{1}{(x^2-4xa_\eta)^{1/2}}
\left[4|h_1|^4\left\{
\frac{a_1x(x-4a_\eta)}{a_1x+(a_1-a_\eta)^2}\right.\right. \nonumber \\
&+&\left.\left.
\frac{a_1(x^2-4xa_\eta)^{1/2}}{x+2a_1-2a_\eta}
\ln\left(\frac{x+(x^2-4xa_\eta)^{1/2}+2a_1-2a_\eta}
{x-(x^2-4xa_\eta)^{1/2}+2a_1-2a_\eta}\right)\right\}
\right.\nonumber\\
&+&\left.
9|h_3|^4\left\{
\frac{a_3x(x-4a_\eta)}{a_3x+(a_3-a_\eta)^2}  \right.\right. \nonumber \\
&+&\left.\left.
\frac{a_3(x^2-4xa_\eta)^{1/2}}{x+2a_3-2a_\eta}
\ln\left(\frac{x+(x^2-4xa_\eta)^{1/2}+2a_3-2a_\eta}
{x-(x^2-4xa_\eta)^{1/2}+2a_3-2a_\eta}\right)\right\}\right] 
\label{lv2}
\end{eqnarray}
for $\ell_\alpha\ell_\beta \rightarrow \eta\eta$.
Here we note that cross terms are cancelled because of the 
assumed flavor structure (\ref{yukawa}). 
Although there are other lepton number violating processes 
$N_{R_i}N_{R_j}\rightarrow \ell_\alpha\ell_\beta$ induced through the $\eta$
exchange, they can be safely neglected in the analysis due to the
additional suppression caused by the smallness of $\lambda_5$.
 
As the lepton number conserving scattering processes which contribute to
determine the number density of $N_{R_2}$, we have
\begin{eqnarray}
\hat\sigma^{(2)}_{N_1N_2}(x)&=&\frac{1}{4\pi}\left[2(2+C^2)|h_1|^2|h_2|^2
\frac{\sqrt{\lambda_{12}}}{x}\left(1+
\frac{(a_1-a_\eta)(a_2-a_\eta)}{(a_1-a_\eta)(a_2-a_\eta)+xa_\eta}
\right.\right.\nonumber \\
&+&\left.\left.\frac{a_1+a_2-2a_\eta}{x}L_{12}\right)
- 4{\rm Re}[(h_1^\ast h_2)^2]
\frac{2\sqrt{a_1a_2}L_{12}}{x-a_1-a_2+2a_\eta}\right], \nonumber\\
\hat\sigma^{(2)}_{N_3N_2}(x)&=&\frac{1}{4\pi}\left[3(2+C^2)|h_3|^2|h_2|^2
\frac{\sqrt{\lambda_{32}}}{x}\left(1+
\frac{(a_3-a_\eta)(a_2-a_\eta)}
{(a_3-a_\eta)(a_2-a_\eta)+ xa_\eta}
\right.\right. \nonumber\\
&+&\left.\left.\frac{a_3+a_2-2a_\eta}{x}L_{32}\right)
- C^2{\rm Re}[(h_3^\ast h_2)^2]
\frac{2\sqrt{a_3a_2}L_{32}}{x-a_3-a_2+2a_\eta}\right]
\label{nlv1}
\end{eqnarray}
for $N_{R_i}N_{R_2} \rightarrow \ell_\alpha\bar\ell_\beta$ which are 
induced through the $\eta$ exchange and also
\begin{eqnarray}
\hat\sigma^{(3)}_{N_1N_2}(x)&=&
\frac{1}{\pi}\left[|h_1|^2|h_2|^2
\left\{\frac{(x-4a_\eta)^{1/2}}{x^{1/2}}
\frac{\sqrt{\lambda_{12}}}{x}
\Big(-2  \right.\right. \nonumber \\
&+&\left.\frac{4a_\eta(a_1- a_2)^2}
{(a_\eta-a_1)(a_\eta-a_2)x +(a_1-a_2)^2a_\eta}\Big) 
+\left(1-2\frac{a_\eta}{x}\right)
L_{12}^\prime\right\}  \nonumber\\
&-&\left.{\rm Re}[(h_1^\ast h_2)^2]
\left(\frac{\sqrt{\lambda_{12}}}{x} +
\frac{2(a_\eta^2-a_1a_2)L_{12}^\prime}
{(x^2-4xa_\eta)^{1/2}(x-a_1-a_2-2a_\eta)}
\right)\right], \nonumber \\
\hat\sigma^{(3)}_{N_3N_2}(x)&=&\frac{C^2}{4\pi}\left[|h_3|^2|h_2|^2
\left\{\frac{(x-4a_\eta)^{1/2}}{x^{1/2}}
\frac{\sqrt{\lambda_{32}}}{x}
\Big(-2  \right.\right. \nonumber \\
&+&\left.\frac{4a_\eta(a_3- a_2)^2}
{(a_\eta-a_3)(a_\eta-a_2)x +(a_3-a_2)^2a_\eta}\Big) 
+\left(1-2\frac{a_\eta}{x}\right)
L_{12}^\prime\right\}  \nonumber\\
&-&\left.{\rm Re}[(h_3^\ast h_2)^2]
\left(\frac{\sqrt{\lambda_{32}}}{x} +
\frac{2(a_\eta^2-a_3a_2)L_{32}^\prime}
{(x^2-4xa_\eta)^{1/2}(x-a_3-a_2-2a_\eta)}
\right)\right]
\label{nlv2}
\end{eqnarray}
 for $N_{R_i}N_{R_2} \rightarrow \eta\eta^\dagger$ which are 
induced through the $\ell_\alpha$ exchange.
It may be useful to note that the cross terms in these reduced cross
sections become zero if the maximum CP phases are assumed
as $\sin 2(\varphi_2-\varphi_{1,3})=1$.

\newpage
\bibliographystyle{unsrt}

\end{document}